\documentclass[12pt]{article}

\usepackage{times}
\usepackage{epsfig,amsmath}
\usepackage{subfigure}
\usepackage{graphicx}
\usepackage{color}
\usepackage{epstopdf}

\newenvironment{sciabstract}{%
\begin{quote} \bf}
{\end{quote}}

\newcounter{lastnote}

\title{Topologically Protected Quantum Entanglement}

\author
{Yao Wang,$^{1,2,3,\dagger}$ Yong-Heng Lu,$^{1,3,\dagger}$ Jun Gao,$^{1,2,3,\dagger}$ Ruo-Jing Ren,$^{1,3}$\\
Yi-Jun Chang,$^{1,3}$ Zhi-Qiang Jiao,$^{1,3}$, Zhe-Yong Zhang,$^{1,3}$ Xian-Min Jin$^{1,3,4,\ast}$\\
\\
\normalsize{$^1$School of Physics and Astronomy, Shanghai Jiao Tong University, Shanghai 200240, China}\\
\normalsize{$^2$Institute for Quantum Science and Engineering and Department of Physics,}\\
\normalsize{Southern University of Science and Technology, Shenzhen 518055, China}\\
\normalsize{$^3$Synergetic Innovation Center of Quantum Information and Quantum Physics,}\\
\normalsize{University of Science and Technology of China, Hefei, Anhui 230026, China}\\
\normalsize{$^4$Institute of Natural Sciences, Shanghai Jiao Tong University, Shanghai 200240, China}\\
\normalsize{$^\dagger$These authors contributed equally to this work}\\
\normalsize{$^\ast$E-mail: xianmin.jin@sjtu.edu.cn}\\
}

\date{}%\today

\begin{document}
\baselineskip24pt

\maketitle

\begin{sciabstract}
Quantum entanglement, as the strictly non-classical phenomena, is the kernel of quantum computing and quantum simulation, and has been widely applied ranging from fundamental tests of quantum physics to quantum information processing. The decoherence of quantum states restricts the capability of building quantum simulators and quantum computers in a scalable fashion. Meanwhile, the topological phase is found inherently capable of protecting physical fields from unavoidable fabrication-induced disorder, which inspires the potential application of topological protection on quantum states. Here, we present the first experimental demonstration of topologically protected quantum polarization entangled states on a photonic chip. The process tomography shows that quantum entanglement can be well preserved by the boundary states even when the chip material substantially introduces relative polarization rotation in phase space. Our work links topology, material and quantum physics, opening the door to wide applications of topological enhancement in genuine quantum regime.\\
\end{sciabstract}
%\paragraph*{Introduction.}
\subsection*{Introduction.}
Quantum correlation differing from classical phenomena lies in multiple indistinguishable particles and such a quantum behavior can provide a computational advantage. For example, the multi-particle quantum walk is able to realize the universal computation efficiently~\cite{g2_QW,Q_multi}. In contrast, the single-particle quantum walks~\cite{QW_1D,QW_2D}, algorithm computing in special structure~\cite{FH} and boson sampling~\cite{BS_1,BS_2,BS_3,BS_4} lacking correlation inside can be mapped to classical wave phenomena precisely, which limits the advantage of uniquely quantum mechanical behavior. Thus, the quantum correlation feature plays a key role in constructing quantum simulators and quantum computers.
 
Beyond the quantum correlation, quantum entanglement as the heart of quantum mechanics highlights the nonseparability and nonlocality of quantum mechanics and is the crucial kernel in quantum communication and computation. The entangled state has been created experimentally in different physical systems~\cite{Entangle_0,Entangle_1,Entangle_3}. Wherein, the photonic entanglement is the promising platform for quantum information processing by reason of that the photon is easy to be created and detected without extremely demanding experimental environment. Among kinds of photonic entanglement including momentum, time, polarization, orbital angular momentum and hyper-entanglement, the polarization entanglement is widely studied and is practically realized in the fields varying from quantum cryptography to quantum communication~\cite{Entangle_2}.

The obstacle in constructing quantum simulators and quantum computers is the unrealized low-decoherence regime of system, which is the constant pursuit in the field. Many efforts have been undertaken including quantum error correction~\cite{QEC_exp,QEC_rev} and quantum topological computing~\cite{Top_Q_com,Top_Q_com_book} but these attempts still haven't been experimentally realized yet. Recently, an alternative way of protecting quantumness using topological phase is proposed and demonstrated experimentally.

Topological photonics aims to topologically protect photons from the inevitable dissipation from fabrication-induced disorder~\cite{Topo_review_1,Topo_review_2}. The topological phase displays an extraordinary robustness to smooth changes in material parameters or disorder and then inherently owns the capability of protecting physical fields~\cite{review1,review2}. Recently, the experimental demonstration on the protection of topological phase on single-photon quantum feature from the diffusion and ambient environment is reported~\cite{Qutop_1photon}. Meanwhile, the protection capacity of the topological phase on generating correlated photon pairs on ring resonators~\cite{Qutop_source}, topologically protected quantum interference~\cite{Qutop_interfere,Qutop_interface} and topologically protected two-photon quantum state~\cite{Qutop_2photon_1,Qutop_2photon_2} are also experimentally demonstrated, which inspires more exploration in `quantum topological photonics', a crossover between topological photonics and quantum information.

Here, we step forward in quantum topological photonics and experimentally demonstrate the topologically protected quantum polarization entangled state on a photonic chip. We firstly further show that the quantum correlation is well preserved in the topological boundary state but decohere exponentially in trivial bulk state as function of evolution distance. The quantum polarization entangled state is further shown to be preserved whether one of them or both two photons of the entangled pair transmits in the topologically protected states. The process tomography shows that quantum entanglement can be well preserved by the boundary states even when the chip material substantially introduces relative polarization rotation in phase space. Our results demonstrate the topological phase provides the protection on quantumness and can be implemented in future quantum information processing.

%\paragraph*{Experimental implement and results.}
\subsection*{Experimental implement and results.}
\begin{figure}[htbp]
	\centering
	\includegraphics[width=1.0\linewidth]{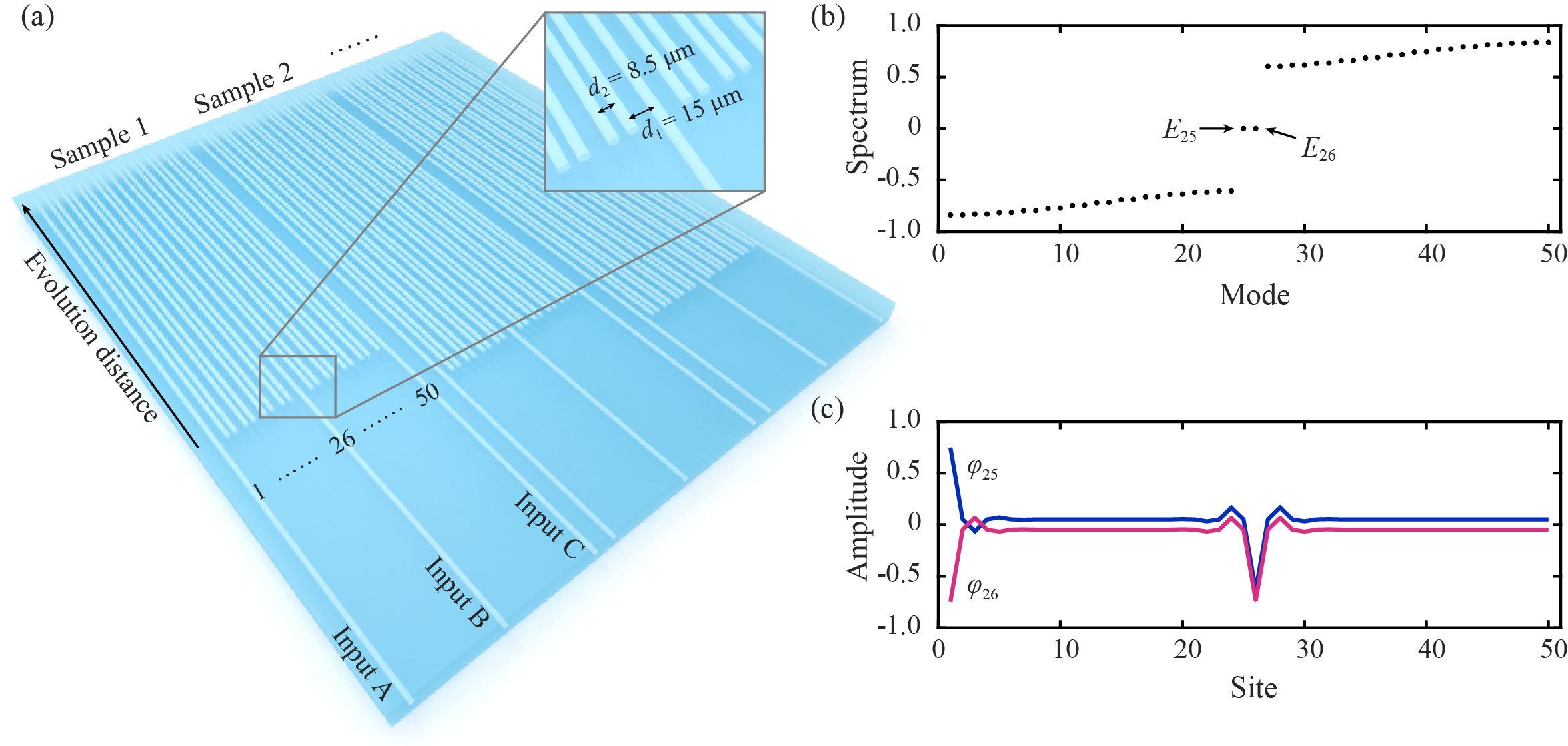}
	\caption{\textbf{Schematic of topological photonic chip.} \textbf{(a)}Seven lattices with different evolution distance varying from 20 mm to 140 mm in photonic chip. The short and long separations between adjacent waveguides is adopted as 8.5 mm and 15 mm respectively . Insert I shows the details of input B. \textbf{(b)} The spectrum of the photonic lattice. $E_{25}$ and $E_{26}$ are two zero-energy modes inside the band gap and are topologically protected. \textbf{(c)} The amplitude of mode $E_{25}$ and $E_{26}$. The photon is mainly confined on the boundary (1st site) and the defect (26th site) of lattice under the norm of the topologically protected zero-energy states.}
	\label{f1}
\end{figure}

In our experiment, we implement the topological lattices of the Su-Schrieffer-Heeger model on a photonic chip using the femtosecond direct laser writing technique~\cite{fabri_1,PIT_Gap} (see Fig.\ref{f1}(a) and Methods for details). The constructed one-dimensional lattices with alternating weak and strong couplings (corresponding separation distances are $d_1=15\ \mu m$ and $d_2=8.5\ \mu m$ respectively) are the conceptually simplest system possessing topological trivial and nontrivial state~\cite{SSH}. The lattice of waveguide array contains 50 sites and the evolution distance varies from 20 mm to 140 mm with step of 20 mm. The spectrum of the lattice is shown in the Fig.\ref{f1}(b), there are two zero-energy modes $E_{25}$ and $E_{26}$ inside the band gap. We further show the spatial distribution of these two modes in Fig.\ref{f1}(c), which can be defined by
\begin{equation}
D_n(E)=\sum_{m}\delta(E-E_m)|\varphi_n^{(m)}|^2,
\end{equation}
where $E_m$ is the energy of the $m$th eigenstate $\varphi_n^{(m)}$. As the result shown, the photons will be confined on the boundary (1st site) or the defect (26th site) of lattice under the norm of the topologically protected zero-energy states if we inject the photon to the lattice from the 1st or 26th site. Thus, the 1st, 26th and 50th site is the topologically protected edge, defect and trivial edge of the lattice and we set these three site as input A, B and C respectively.

The quantum correlated signal and idler photon are injected into the lattice from input A and B respectively, and they are localized in the excited waveguides under the norm and protection of topological phase. We detect the photons from the excited waveguides at output facet of photonic chip and record the coincidence of the photon pairs for 700 seconds using a home-made FPGA. To quantify the performance of two-photon quantum correlation after the photons coming from the lattices, we calculate the cross-correlation function as~\cite{Q_optics}
\begin{equation}
g^{(2)}_{s\text{-}i} = p_{s\text{-}i}/p_sp_i,
\end{equation}
where $p_{s\text{-}i}$ is the coincidence probability of signal and idler photons, $p_s$ ($p_i$) is the detection probability of signal (idler) photon. As comparison, we inject two photons into the lattice from input B and C as the trivial case.

\begin{figure}[htbp]
	\centering
	\includegraphics[width=0.85\linewidth]{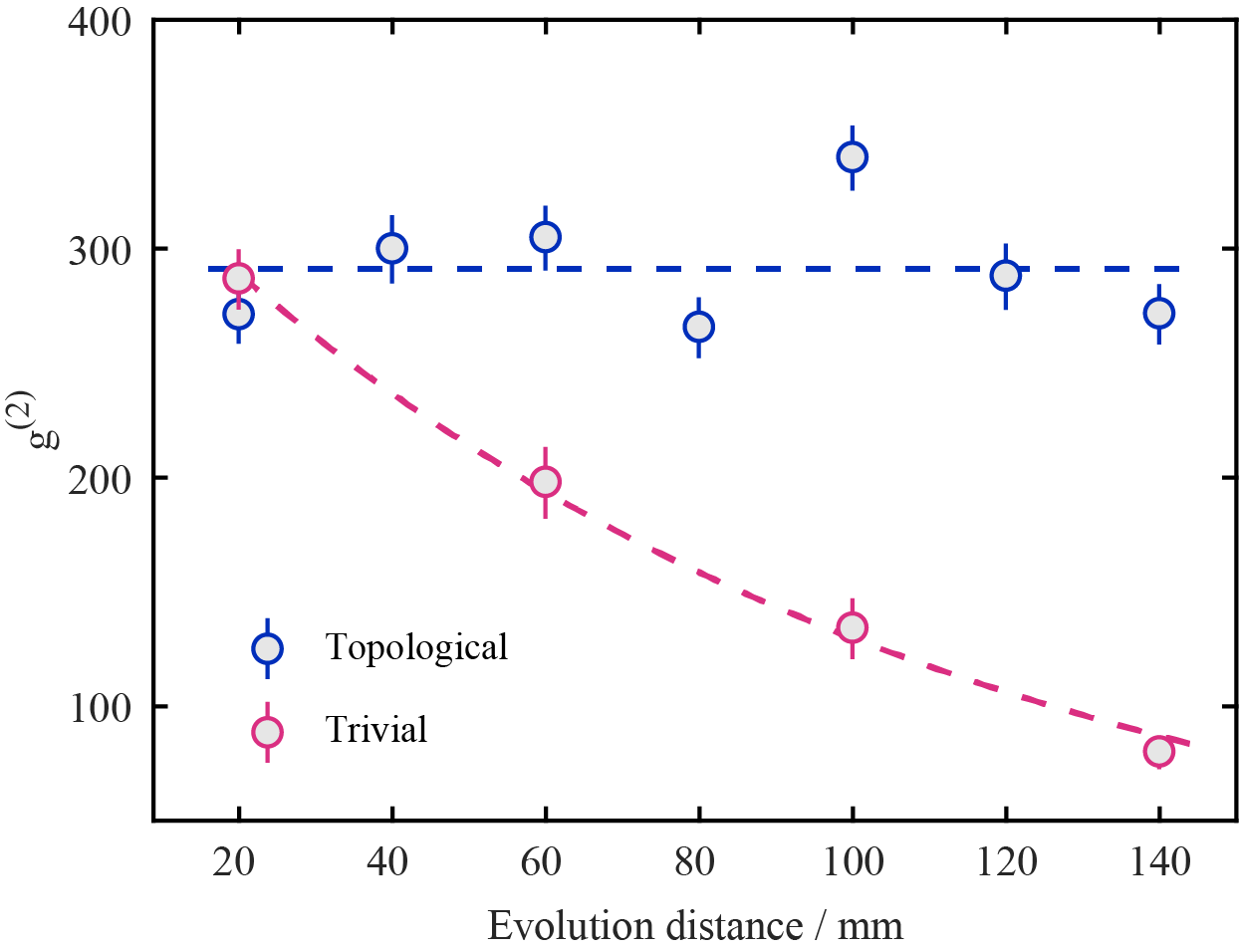}
	\caption{\textbf{Measured topologically protected two-photon state.} The two-photon state is measured for the topological cases of $z=20$ mm to 140 mm in a step of 20 mm and the trivial cases of $z=20$ mm to 140 mm in a step of 40 mm. The quantum correlation is well preserved in topological state but undergoes exponential decoherence in trivial state}
	\label{f2}
\end{figure}

The results show that the $g^{(2)}_{s\text{-}i}$ is preserved in high level up to 300 and has no drop with the increase of evolution distance for the two-photon state in topological edge state, while the there is an exponential decay for the value of $g^{(2)}_{s\text{-}i}$ with the evolution for the two-photon state in the trivial bulk case [see Fig.\ref{f2}]. The average value of $g^{(2)}_{s\text{-}i}$ of topological case is 291.34. We fit the result of trivial case with exponential function, and the curve function is $g^{(2)}_{s\text{-}i}=353.02e^{-0.01z}$, where $z$ is the evolution distance in millimeter.

We demonstrate the topological protection of two-photon quantum state in different evolution distance beyond previous work. Though the feature of quantum correlation plays a key role in quantum computing and simulation, the quantum entanglement lies in the heart of quantum mechanics are the crucial kernel in quantum communication and computation. The next step forward is to investigate whether it is accessible to extend promised topological protection into quantum entanglement regime.

\begin{figure}[htbp]
	\centering
	\includegraphics[width=1.0\linewidth]{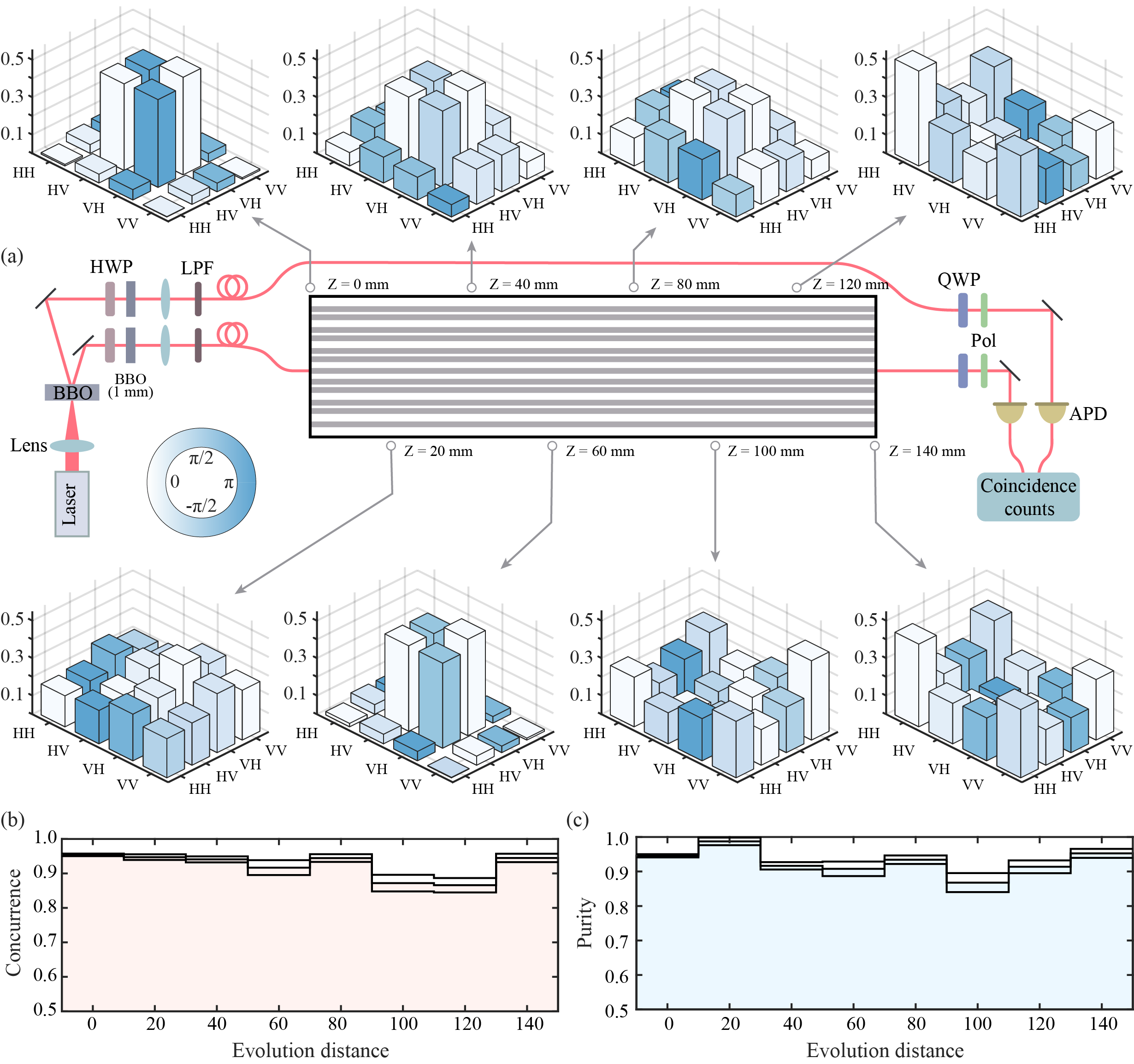}
	\caption{\textbf{Measured tomography of topologically protected entangled state.} \textbf{(a)} The experimental setup and result. Two-photon polarization entangled state is generated by the process of type-II spontaneous parametric down-conversion. One of the two entangled photons is injected into the lattices from input B, and the state tomography is conducted for different evolution distance. The modulus and argument of the matrix elements are represented by the height and color of the bars respectively. \textbf{(b-c)} The values of concurrence (b) and purity (c) don't change much with the increase of evolution distance beyond 90\%. The results imply that the entangled state is well protected. BBO: beat-barium-borate, HWP: half-wave plate, QWP: quarter wave plate, LPF: long-pass filter, Pol: polarizer, APD: avalanche photodiode.}
		\label{f3}
\end{figure}

As shown in Fig.\ref{f3}, we prepare an entangled state
\begin{equation}
	|\psi^-\rangle = \frac{|HV\rangle - |VH\rangle}{\sqrt2},
\end{equation}
and then inject one of the photon into the lattice from input B, and carry out quantum state tomography to reconstruct the density matrix of the output states for different evolution distance. To quantify the performance of entangled state out of the photonic chip, we calculate the concurrence of the measured two-qubit entangled state, which is given by
\begin{equation}
	C = max\{0,\Gamma\},
\end{equation}
where $\Gamma=\sqrt{\lambda_1}-\sqrt{\lambda_2}-\sqrt{\lambda_3}-\sqrt{\lambda_4}$. The quantities $\lambda_j$ are the eigenvalues in decreasing order of the matrix $\rho(\sigma_x\otimes\sigma_y)\rho^\ast(\sigma_y\otimes\sigma_x)$, where $\sigma_y$ is the second Pauli matrix and the variable $\rho^\ast$ corresponds to the complex conjugate of the density matrix of the state $\rho$ in the canonical basis $\{|HH\rangle,|HV\rangle,|VH\rangle,|VV\rangle\}$. As shown in Fig.\ref{f3}(b), the calculated values of concurrence keep in high beyond 90\%, which implies that the two-qubit entangled state is preserved well by the topological edge state.

As a quantity describing the quantum state, the purity gives the information on how much a state is mixed, which is defined by
\begin{equation}
	\gamma\equiv\mbox{tr}(\rho^{2}).
\end{equation}
If the value $\gamma$ equals to 1, then the state is pure. The calculated purity of the measured states are shown in Fig.\ref{f3}(c), similar to the result of concurrence, the values of purity don't change much with the increase of evolution distance beyond 90\%, which implies the state keeps in a well pure state while it transmits in the topological defect state.

\begin{figure}[htbp]
	\centering
	\includegraphics[width=1.0\linewidth]{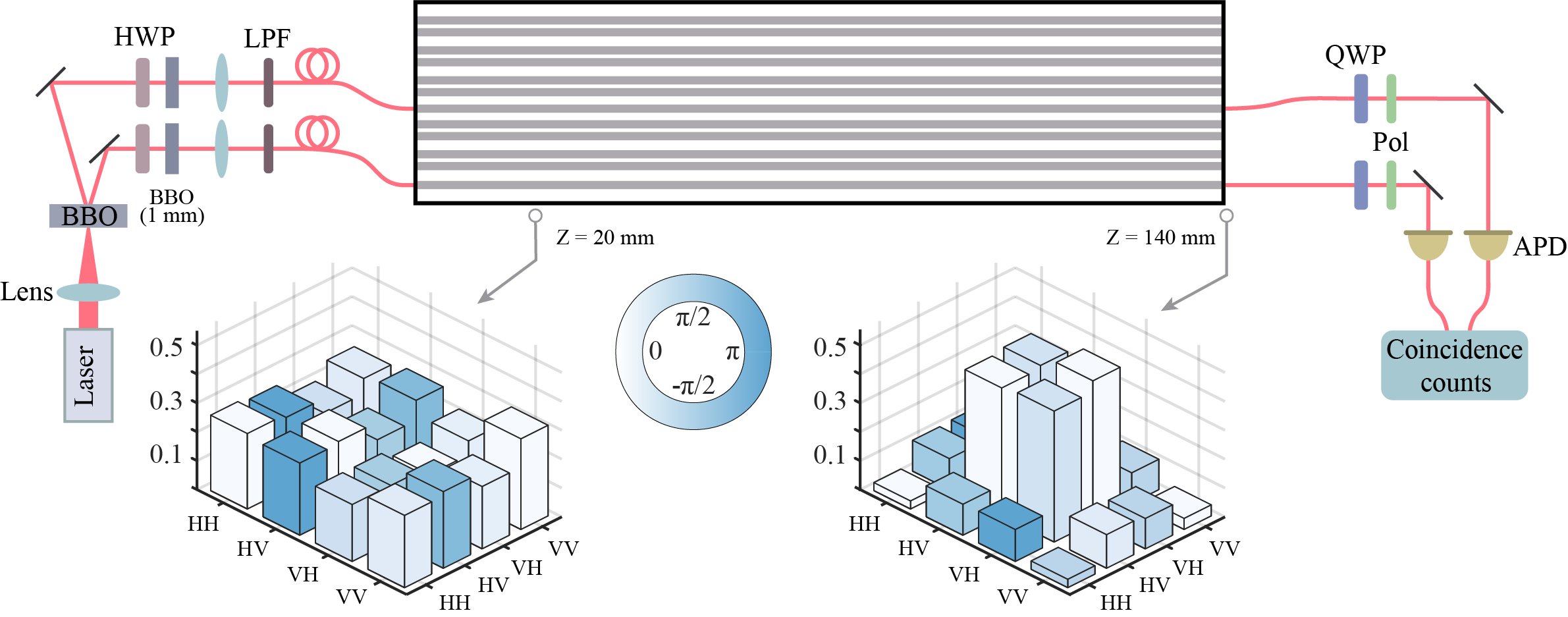}
	\caption{\textbf{Measured density matrix of topologically protected entangled state.} Both two entangled photons are injected into the lattices from input A and B. The tomography for the case of evolution distance $z=20$ mm and $z=140$ mm are measured. The corresponding concurrence is $0.88\pm0.02$ and $0.95\pm0.02$ respectively. The value of purity is $0.89\pm0.02$ for $z=20$ mm and $0.94\pm0.03$ for $z=140$ mm. The modulus and argument of the matrix elements are represented by the height and color of the bars respectively.}
	\label{f4}
\end{figure}

Based on the above result, we may get a quick guess that the entangled state can also be preserved when both the two photons transmit in topologically protected states. To further investigate the performance of the quantum entangled state in topological lattice, we inject the two photons into the photonic chip from input A and B respectively. The results of the reconstructed entangled density matrix in photonic lattices of $z=20$ mm and 140 mm are shown in Fig.\ref{f4}, and the corresponding concurrence is $0.88\pm0.02$ and $0.95\pm0.02$ respectively. The value of purity is $0.89\pm0.02$ for $z=20$ mm and $0.94\pm0.03$ for $z=140$ mm. The results verify our assumptions, and the entangled state is also well preserved when the two photons transmit in the topological edge state.

Comparing the measured tomography results with the source [see the result of $z=0$ mm in Fig.\ref{f2}(a)], some quantities have changed during the entangled photons transmission in the topological photonic chip, though the entanglement is well preserved. The process tomography as a measurement method can reveal the details of the topological channel and specify the change of the entangled state in the lattices. 

\begin{figure}[htbp]
	\centering
	\includegraphics[width=1.0\linewidth]{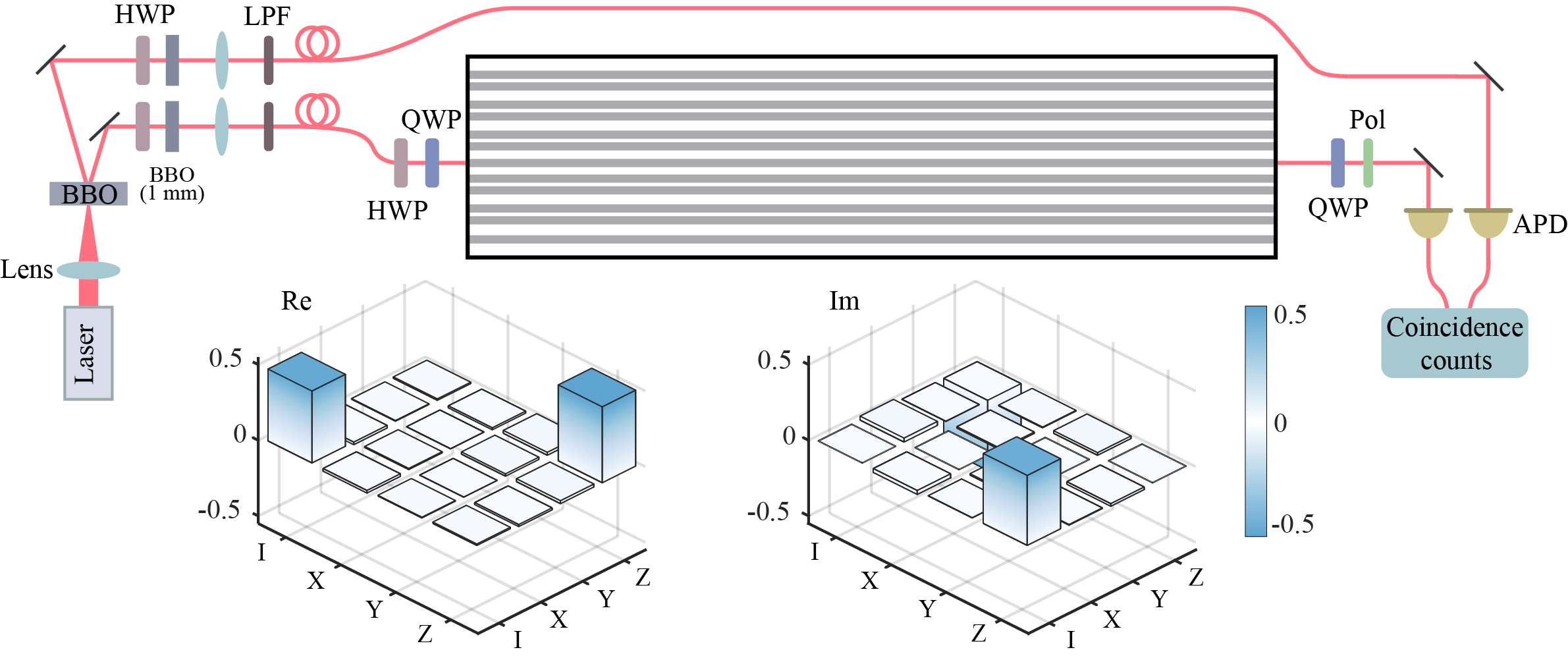}
	\caption{\textbf{Process tomography result.} The photon is injected into the lattice from the input B. A HWP and a QWP are placed before the photonic chip while a QWP and a Pol are placed behind the chip. The real and imaginary parts of the process matrix elements are represented in two independent figures.}
	\label{f5}
\end{figure}

Taking the case of exciting lattice in input A and B with the evolution distance $z = 140$ mm as an example, we conduct the process tomography and show the calculated result applying the convex optimization~\cite{CO} in Fig.\ref{f5}. The obtained process matrix $\chi$ can describe the true operations of the quantum channel. From the matrix shown in Fig.\ref{f5}, we can derive that
\begin{eqnarray}
	\chi&=&\frac{1}{2}I\rho I-\frac{i}{2}I\rho Z+\frac{i}{2}Z\rho I+\frac{1}{2}Z\rho Z, \\ \nonumber
	&=&(\frac{I+Z}{2}-i\frac{I-Z}{2})\rho(\frac{I+Z}{2}+i\frac{I-Z}{2}), \\ \nonumber
	&=&S\rho S^\dagger,
\end{eqnarray}
where $S=\begin{bmatrix} 1&0\\0&-i \end{bmatrix}$ is the phase gate and it is unitary because $SS^\dagger=I$. The channel only induce some phase change to the corresponding input state, and will not corrupt the original information. In other words, the quantum entanglement can be well preserved by the topologically protected states even when the chip material substantially introduces relative polarization rotation in phase space.

One more point should be noted, we demonstrate the performance of topologically protected quantum correlation and entanglement in different evolution distance varying from 20 mm to 140 mm, which are realized in different lattice. Seven different lattice samples are fabricated for the measurement in our experiment. The disorder induced by fabrication in each lattice is inevitable though we have locked or optimized all the parameters of the system and the fabrication environment during femtosecond laser direct-write process. However, the performance of quantum correlation and entanglement in experimental result is not influenced by the fabrication-induced disorder, such behaviors are owing to the distinguishing feature of topological protected quantum state.

%\paragraph*{Discussion and Conclusion.}
\subsection*{Discussion and Conclusion.}
In conclusion, we experimentally demonstrate the topologically protected polarization entangled state on a photonic chip. We have found that quantum correlation is still well preserved in the topological boundary state after long evolution distance up to 140 mm but the correlated state suffers the exponential decoherence in trivial bulk state as function of evolution distance. The topologically protected quantum entangled state keeps in high concurrence and purity beyond 90\% and don't change much with the increase of evolution distance on the topological photonic chip regardless of the inevitable fabrication-induced disorder. We further demonstrate the details of the transmission, the relative polarization rotation in phase space induced by the chip material while the entanglement is preserved, by conducting quantum process tomography.

Our results extend the protection mechanism of topological phases into quantum entanglement regime, representing an emerging and alternative way of protecting quantumness and can be implemented in future quantum information processing. The demonstrated key elements, including integrated structures and topological protected entangled state, can enrich the emerging field of `quantum topological photonics', a crossover between topological photonics and quantum information. The prompt questions, though remain open, will be the behavior of topological protected multi-photon entanglement dynamics in higher dimensional structures.\\

%\paragraph*{Acknowledgments.}
\subsection*{Acknowledgments.}
The authors thank Roberto Osellame and Jian-Wei Pan for helpful discussions. This research is supported by National Key R\&D Program of China (2017YFA0303700), National Natural Science Foundation of China (NSFC) (61734005, 11761141014, 11690033), Science and Technology Commission of Shanghai Municipality (STCSM) (15QA1402200, 16JC1400405, 17JC1400403), Shanghai Municipal Education Commission (16SG09, 2017-01-07-00-02-E00049), X.-M.J. acknowledges support from the National Young 1000 Talents Plan.\\

\subsection*{Methods}
\paragraph*{Fabrication and measurement of the lattices on a photonic chip:} We fabricate the samples in borosilicate glass substrate (refractive index $n_0=1.514$ for writing laser at a wavelength of 513nm) using the laser system operating at a repetition rate of 1 MHz and a pulse duration of 290fs. The light  is focused inside the sample with a 50X microscope objective (NA=0.50) after being reshaped with a cylindrical lens. We continuously move the substrates using a high-precision three-axis translation stage with a constant velocity of 15 mm/s to create the lattices by the laser-induced refractive index increase.

In the experiment, we inject the photons into the input waveguides in the photonic chip using a 20X objective lens. After a total propagation distance through the lattice structures, the outgoing photons are first collimated with a 10X microscope objective, then detected and analyzed by a combination of wave plates and polarizers.

\paragraph*{Generation of entangled state:} The two-photon polarization entangled state is generated by the process of type-II spontaneous parametric down-conversion. We focuses a 405 nm UV laser on a 2 mm type-II degenerate noncolinear cut BBO (beat-barium-borate) to generate the polarization-entangled singlet state $|\psi^-\rangle$. Especially, we compensate the birefringence induced spatial and temporal distinguishability using the combination of a half-wave plate (HWP) and a BBO with 1 mm thick. Two band pass filters centered at the desired wavelength of the down-converted photons are employed behind the BBO. 

\paragraph*{Quantum state tomography:} We reconstruct and analyze the quantum state by projecting the output state onto quorum states (H, V, D and R) using a section composed of a quarter-wave plate (QWP) and a polarizer (Pol). The QWP and Pol are placed after the photonic chip and before the single-photon detector to sequentially perform the projections to particular polarization states. The two photons are detected after be projected into quorum states respectively, sequentially, there are 16 performed measurements for each quantum state tomography procedure, which can be represented by the tensor products
\begin{equation}
	\{M\} = \{H, V, D, R\}\otimes\{H, V, D, R\}.
\end{equation}
We firstly project one of the photons to polarization H state and then fix it, which is followed by projecting the other photon in polarization H, V, D and R states successively. The residual measurement of the first photon in V, D, R basis can be repeated by the same procedure. Then the result of the measurement can be obtained by counting and recording the coincidence of the photon pairs. A max likelihood optimization is employed to guarantee the physical feature of the reconstructed density matrix. Thus, the precise reconstruction of the output states with the whole density matrix is completed.

\paragraph*{Quantum process tomography:} We analyze the whole changes and transformation of the entanglement state in the topological protected edge channel using the standard quantum process tomography, which is equivalent to quantum state tomography performed on a larger parameter space and similar to an on-chip interferometry network or a quantum circuit processing qubits in a quantum register. The measured process matrix $\chi$ can be expressed on the basis of $\widetilde{E}_m$ ($\widetilde{E}_1=I$ is the identity operation, $\widetilde{E}_2=X$, $\widetilde{E}_3=Y$ and $\widetilde{E}_4=Z$ represent the three Pauli operators) and expressed as
\begin{equation}
\varepsilon = \sum_{mn}\chi_{mn}\widetilde{E}_m\rho\widetilde{E}_m^\dagger.
\end{equation}
The process matrix $\chi$ can completely and uniquely describes the process $\varepsilon$ and can be reconstructed by experimental tomographic measurements.

As for the experimental tomographic measurement, the standard quantum process tomography is composed of input state preparation and output state tomography. We use the combination of half-wave plate (HWP) and QWP to prepare different initial states. One should be noted that the quantum process tomography is implemented with one photon in lattice and the other one behaves as the trigger.The computational basis vectors of the input state preparation that we choose are H, V, D and R and the computational basis vectors of the output state tomography that we choose are H, V, D, A, R and L. Thus the standard quantum process tomography procedure consists of twenty-four sequentially performed measurements, which can be represented by the tensor products
\begin{equation}
	\{M\} = \{H, V, D, R\}\otimes\{H, V, D, R, A, L\}.
\end{equation}
The operations of measurement is similar to the procedure of quantum state tomography.

\paragraph*{Relationship between process matrix $\chi$ and phase gate $S$:} The obtained process matrix $\chi$ can be described as
\begin{equation}
\begin{split}
\chi&=\frac{1}{2}\begin{bmatrix}
1 & 0 & 0 & i\\
0 & 0 & 0 & 0\\
0 & 0 & 0 & 0\\
-i & 0 & 0 & 1
\end{bmatrix}\\
&= \frac{1}{2}I\rho I - \frac{i}{2}I\rho Z + \frac{i}{2}Z\rho I +  \frac{1}{2}Z\rho Z
\end{split}
\end{equation}
with the basis vector$ \{I, X, Y, Z\}$. Meanwhile the phase gate $S$ is
\begin{equation}
\begin{split}
S&=\frac{I+Z}{2}-i\frac{I-Z}{2}\\
&=\frac{1}{2}\left\lbrace\begin{bmatrix} 1&0\\0&1 \end{bmatrix}+\begin{bmatrix} 1&0\\0&-1 \end{bmatrix}\right\rbrace-\frac{i}{2}\left\lbrace \begin{bmatrix} 1&0\\0&1 \end{bmatrix}-\begin{bmatrix} 1&0\\0&-1 \end{bmatrix}\right\rbrace \\
&=\begin{bmatrix} 1&0\\0&-i \end{bmatrix}.
\end{split}
\end{equation}
Such that, we can get the relationship between them as
\begin{equation}
\begin{split}
S\rho S^\dagger & = (\frac{I+Z}{2}-i\frac{I-Z}{2}) \rho (\frac{I+Z}{2}+i\frac{I-Z}{2}) \\
& = (\frac{I+Z}{2}) \rho (\frac{I+Z}{2})+i(\frac{I+Z}{2})\rho(\frac{I-Z}{2})\\
&\ \ \ \ -i(\frac{I-Z}{2}) \rho (\frac{I+Z}{2})+(\frac{I-Z}{2})\rho(\frac{I-Z}{2})\\
& = (\frac{1}{4}I\rho I + \frac{1}{4}I\rho Z +  \frac{1}{4}Z\rho I +  \frac{1}{4}Z\rho Z)\\
&\ \ \ \ +i(\frac{1}{4}I\rho I - \frac{1}{4}I\rho Z +  \frac{1}{4}Z\rho I -  \frac{1}{4}Z\rho Z)\\
&\ \ \ \ -i(\frac{1}{4}I\rho I + \frac{1}{4}I\rho Z -  \frac{1}{4}Z\rho I -  \frac{1}{4}Z\rho Z)\\
&\ \ \ \ +(\frac{1}{4}I\rho I - \frac{1}{4}I\rho Z -  \frac{1}{4}Z\rho I +  \frac{1}{4}Z\rho Z)\\
& = \frac{1}{2}I\rho I - \frac{i}{2}I\rho Z + \frac{i}{2}Z\rho I +  \frac{1}{2}Z\rho Z\\
& = \chi.
\end{split}
\end{equation}

\clearpage
% If your reference list includes text notes as well as references, include the following line; otherwise, comment it out.
%\renewcommand\refname{References and Notes}

\clearpage

%\noindent {\bf Fig. 1.}

\clearpage

%\noindent {\bf Fig. 1.}

\clearpage

%\noindent {\bf Fig. 2.}

\clearpage

\clearpage

\end{document}